\documentclass[aps,prb,reprint,superscriptaddress,floatfix]{revtex4-1}

\usepackage{amsmath}    
\usepackage{graphicx}   
\usepackage[lofdepth,lotdepth, caption=false]{subfig}
\usepackage{verbatim}   
\usepackage{color}      
\usepackage{hyperref}   
\usepackage{dcolumn}
\usepackage{xcolor}  
\usepackage{tabularx}

\begin{document}
\newcommand{\psit}{(Pb$_{0.5}$Sn$_{0.5}$)$_{1-x}$In$_{x}$Te}
\newcommand{\pst}{Pb$_{0.5}$Sn$_{0.5}$Te}
\newcommand{\degrees}{$^\circ$C}
\newcommand{\Tc}{$T_{c}$}
\newcommand{\at}[2][]{#1|_{#2}}
\newcolumntype{Y}{>{\centering\arraybackslash}X}

\title{Superconductivity induced by In substitution into the topological crystalline insulator \pst}
\author{R.~D.~Zhong}
\affiliation{Condensed Matter Physics and Materials Science Department, Brookhaven National Laboratory, Upton, NY 11973, USA}
\affiliation{Materials Science and Engineering Department, Stony Brook University, Stony Brook, NY 11794, USA}
\author{J.~A.~Schneeloch}
\affiliation{Condensed Matter Physics and Materials Science Department, Brookhaven National Laboratory, Upton, NY 11973, USA}
\affiliation{Department of Physics and Astronomy, Stony Brook University, Stony Brook, NY 11794, USA}
\author{T. S. Liu}
\affiliation{Condensed Matter Physics and Materials Science Department, Brookhaven National Laboratory, Upton, NY 11973, USA}
\affiliation{School of Chemical Engineering and Environment, North University of China, Shanxi 030051, China}
\author{F. E. Camino}
\affiliation{Center for Functional Nanomaterials, Brookhaven National Laboratory, Upton, NY 11973, USA}
\author{J.~M.~Tranquada}
\author{G.~D.~Gu}
\email{ggu@bnl.gov}
\affiliation{Condensed Matter Physics and Materials Science Department, Brookhaven National Laboratory, Upton, NY 11973, USA}

\date{\today} 

\begin{abstract}
Indium substitution turns the topological crystalline insulator (TCI) \pst\ into a possible topological superconductor. To investigate the effect of the indium concentration on the crystal structure and superconducting properties of \psit, we have grown high-quality single crystals using a modified floating-zone method, and have performed systematic studies for indium content in the range $0\leq x\leq 0.35$. We find that the single crystals retain the rock salt structure up to the solubility limit of indium ($x\sim0.30$). Experimental dependences of the superconducting transition temperature (\Tc) and the upper critical magnetic field ($H_{c2}$) on the indium content $x$ have been measured. The maximum \Tc\  is determined to be 4.7 K at $x=0.30$, with $\mu_0H_{c2}(T=0)\approx 5$~T. 
\end{abstract}


\maketitle

Topological insulators (TIs) form a class of materials that is currently creating a surge of research activity because they represent a new quantum state of matter in which the bulk is insulating while the surface presents robust gapless states.\cite{Hasan2010, Qi2011} They are made possible because of two major features: symmetry under time reversal and the spin-orbit interaction\cite{Qi2011, Sato2013} In the case of TIs, narrow-gap semiconductors such as Bi$_{2}$Se$_{3}$ and Bi$_{2}$Te$_{3}$, with an odd number of band inversions, support gapless Dirac-like surface states. In contrast, the narrow-gap IV-VI semiconductors PbTe, PbSe, and SnTe were initially considered to be topologically trivial insulators.\cite{Fu2007} More recently, the notion of ``topological crystalline insulators'' (TCIs) was introduced, extending the topological classification of band structures to take account of certain crystal point group symmetries.\cite{Fu2011}  This general proposal soon lead to the specific prediction that rock-salt-structured SnTe should have robust surface states that are symmetric about \{110\} mirror planes,\cite{Hsieh2012} which was quickly confirmed experimentally.\cite{Tanaka2012}  Additional examples of TCIs have been found in the IV-VI substitutional solid solutions Pb$_{1-z}$Sn$_{z}$Te\cite{Xu2012, Safaei2013} and Pb$_{1-z}$Sn$_{z}$Se.\cite{Dziawa2012, Wojek2013,Pletikosic2014}  

The discovery of TIs and TCIs has also stimulated an experimental search for topological superconductors (TSCs), whose surface states should have the character of Majorana fermions. \cite{Qi2011}  In particular, spectroscopic studies have recently provided evidence for odd-parity pairing and topological superconductivity in In-doped SnTe\cite{Sasaki2012}; another candidate system is Cu$_{x}$Bi$_{2}$Se$_{3}$.\cite{Fu2010,Hor2010,Sasaki2011}  The guiding principle has been to look for superconductivity in low-carrier-density narrow-gap semiconductors with strong spin-orbit coupling and with Fermi surfaces centered around time-reversal-invariant momenta.\cite{Sasaki2012, Hsieh_Fu2012, Fu2010, Michaeli2012}  It should be pointed out that, while one might expect  compositional disorder or impurity scattering to destroy the superconducting order, it is predicted to survive due to an emergent chiral symmetry.\cite{Michaeli2012} Thus, we have been motivated to consider a new TSC candidate: In-substituted Pb$_{1-z}$Sn$_{z}$Te.

Earlier studies on polycrystalline samples of (Pb$_{1-z}$Sn$_{z}$)$_{1-x}$In$_x$Te investigated the dependence of $T_c$ and $H_{c2}$ in the concentration ranges $0.4\le z\le1$ and $0.03\le x\le0.2$.\cite{Bushmarina1991, Parfeniev1995, Parfeniev2001}  The maximum $T_c$ was found for $z=0.5$; however, the solubility limit in these polycrystalline samples was reached by $x\sim0.2$.  The Pb content of $z=0.5$ is the likely region of interest for the TCI state, as, without In, Pb$_{1-z}$Sn$_{z}$Te undergoes a transition from a trivial insulator to a TCI as $z$ increases through 0.35.\cite{Dziawa2012, Dimmock1966, Tanaka2013}  These factors determine our choice of $z=0.5$ for the present investigation.

In this paper, we report the growth of \psit\ single crystals for $0\leq x\leq 0.35$ using the modified floating-zone method.  The crystals have been characterized by scanning electron microscopy, x-ray diffraction, magnetization, and resistivity.  We find the $T_c$ reaches a maximum of 4.7(1)~K at $x=0.30(1)$, which is the approximate solubility limit of In in crystals of this alloy. 

Single crystals with nominal In concentrations of $x_{\rm nom}=0$, 0.1, 0.15, 0.2, 0.3, and 0.4 were grown. For each sample, high purity (99.999\%) elemental forms of Pb, Sn, In, and Te were loaded into double-walled quartz ampoules and sealed under vacuum. The materials were melted at 900\degrees\ in a box furnace, with rocking to achieve homogeneous mixing, and then cooled to room temperature at $10^\circ$C/h. The outer quartz tube was then removed, and the inner tube was mounted in the floating-zone furnace, surrounded by an atmosphere of 1 bar Ar.  The solidified, cylindrical ingot was first pre-melted at a velocity of 200 mm/hr, and then crystal growth was performed at 1.0 mm/hr.  Because the segregation coefficient $k_{\rm sg}$ of In is less than 1, the In contained in the feed material would prefer to remain in the liquid zone, and thus the In concentration gradually grows with time. As a result, we anticipate an In concentration gradient along the length of the as-grown crystal rod, with a typical total length of 150 mm.  To account for this, the ends of each crystal were removed and discarded; the central 100 mm was then cut into 20-mm sections.  For each section used for further study, the In concentration was measured on a piece from each end using energy-dispersive X-ray spectroscopy (EDS).  In each case, these measured $x$ values agreed within the measurement uncertainty.  We use these measured $x$ values throughout the rest of the paper.

\begin{figure}[t]
\begin{center}
\includegraphics[width=\columnwidth]{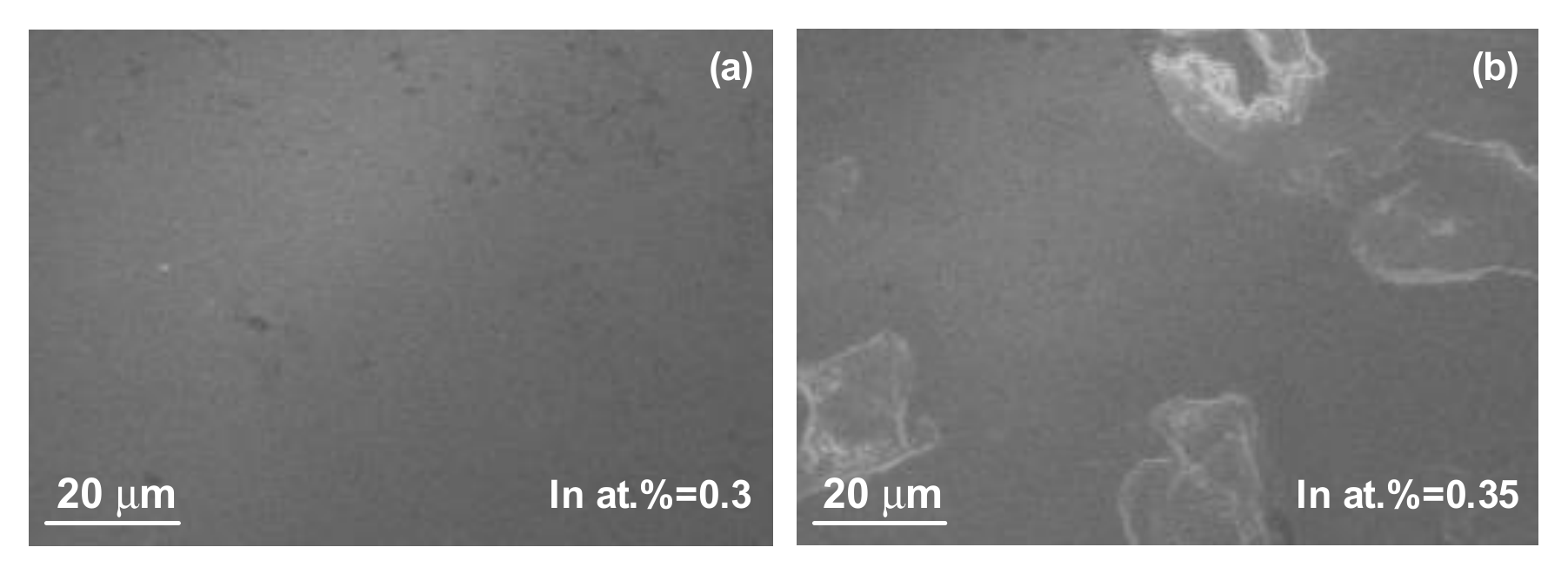}
\caption{\label{fig:SEM} Scanning electron microscope (SEM) images for (a) 30\% and (b) 35\% In substituted \psit\ crystals. The bright regions seen in (b) have a composition consistent with InTe; these are absent from (a).}
\end{center}
\end{figure}

\begin{figure}[t]
\begin{center}
\includegraphics[width=\columnwidth]{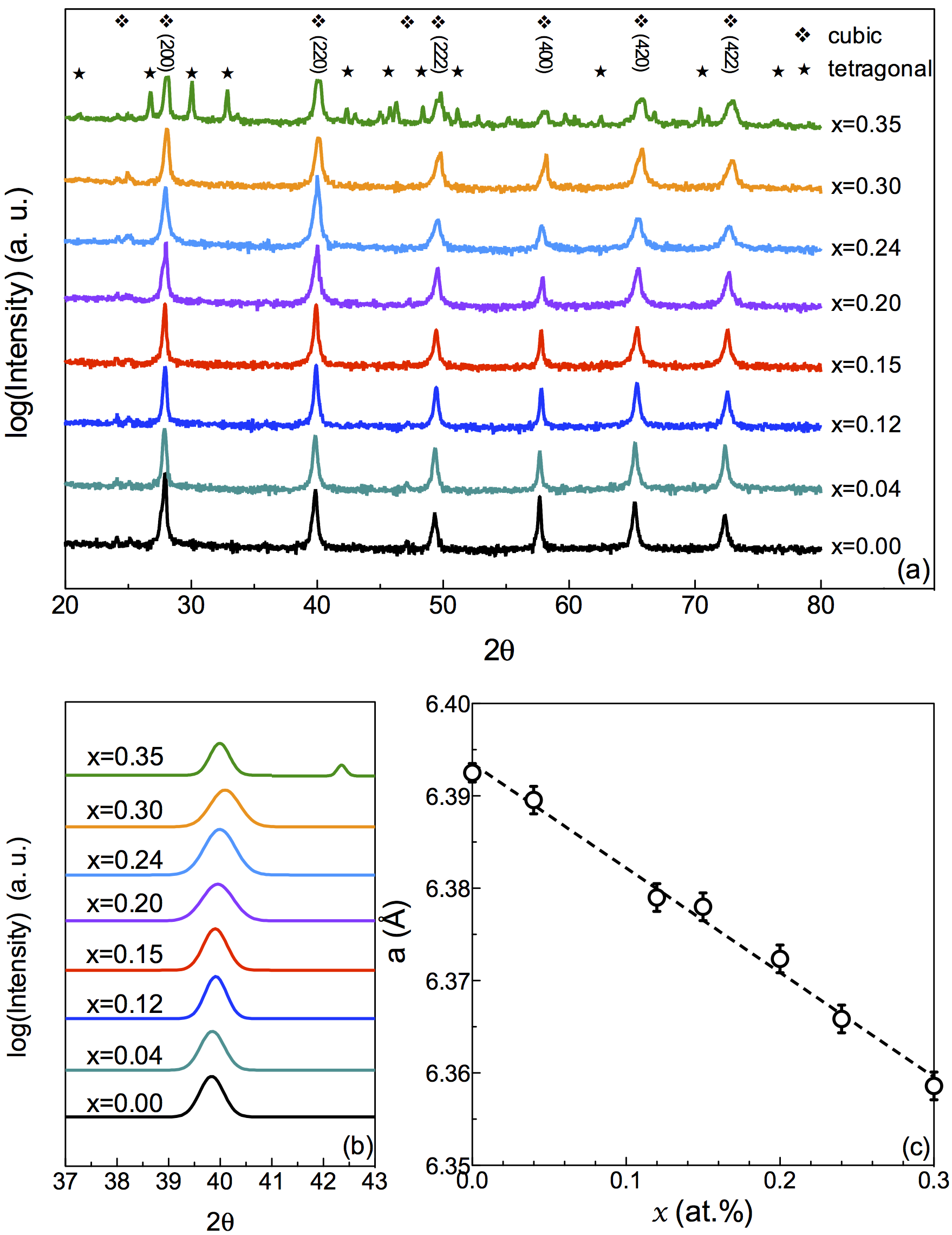}
\caption{\label{fig:XRD} (Color online) (a) X-ray diffraction patterns for the \psit\ single crystals with composition $x=0$--0.35, with intensity plotted on a logarithmic scale. The data have been smoothed using the Savitzky-Golay algorithm, and the $K\alpha_{2}$ component has been removed.\cite{deRooi2014} The Miller indices of major peaks in the cubic phase have been identified. (b) Gaussian fitted curves for the (220) peak in the cubic phase. (c) Lattice parameters $a$ of the cubic phase as a function of the indium concentration $x$.}
\end{center}
\end{figure}

Microstructural and compositional investigations of the crystals were performed using an analytical high resolution scanning electron microscope (SEM) equipped for EDS, model JEOL 7600F, located at the Center for Functional Nanomaterials (CFN).   For each crystal piece characterized, EDS was measured at 10 positions, and the variation in $x$ was generally found to be $<2$\%~of the mean value.  SEM images in Fig.~\hyperref[fig:SEM]{1} show typical microstructures of the \psit\ cleaved surface.  Figure~\hyperref[fig:SEM]{1(a)} is representative for $x\le0.3$, with a dense uniform microstructure having few voids, consistent with single-phase behavior.   In contrast, the $x=0.35$ sample shown in Fig.~\hyperref[fig:SEM]{1(b)} exhibits a secondary phase, confirmed as InTe by EDS analysis, dispersed randomly in the majority phase. 

\begin{figure}[t]
\begin{center}
\includegraphics[width=0.8\columnwidth]{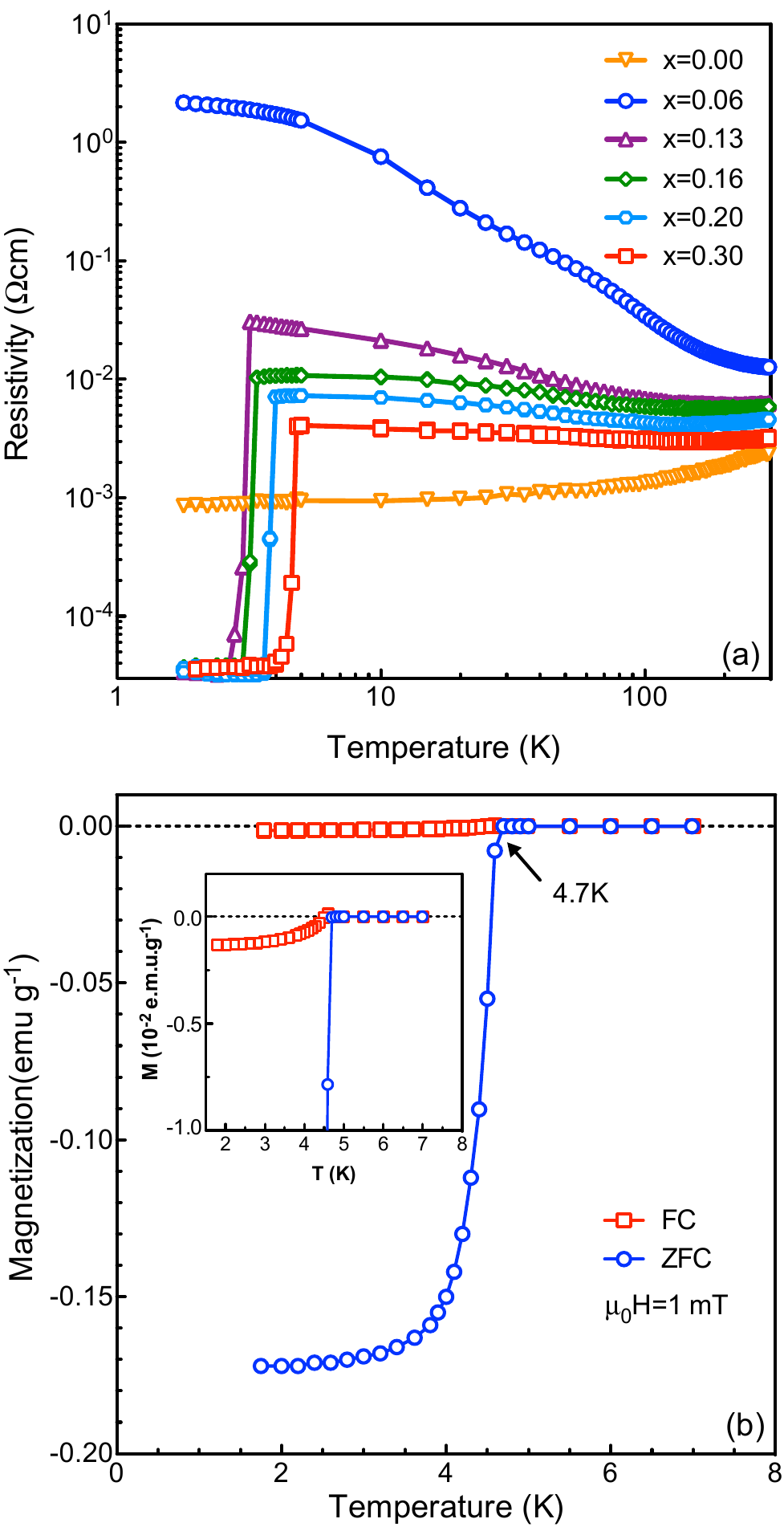}
\caption{\label{fig:MTRT} (Color online)  (a) Temperature dependence of the resistivity for \psit\ single crystals with indium contents $x=0$--0.30.  (b) Temperature dependence of magnetic susceptibility for an optimally doped \psit\ ($x$=0.30) single crystal measured under conditions of field cooling (FC, red squares) and zero-field cooling (ZFC, blue circles) in an applied field of 1~mT at a heating/cooling rate of 0.1 K/min.}
\end{center}
\end{figure}

\begin{figure}
\begin{center}
\includegraphics[width=0.7\columnwidth]{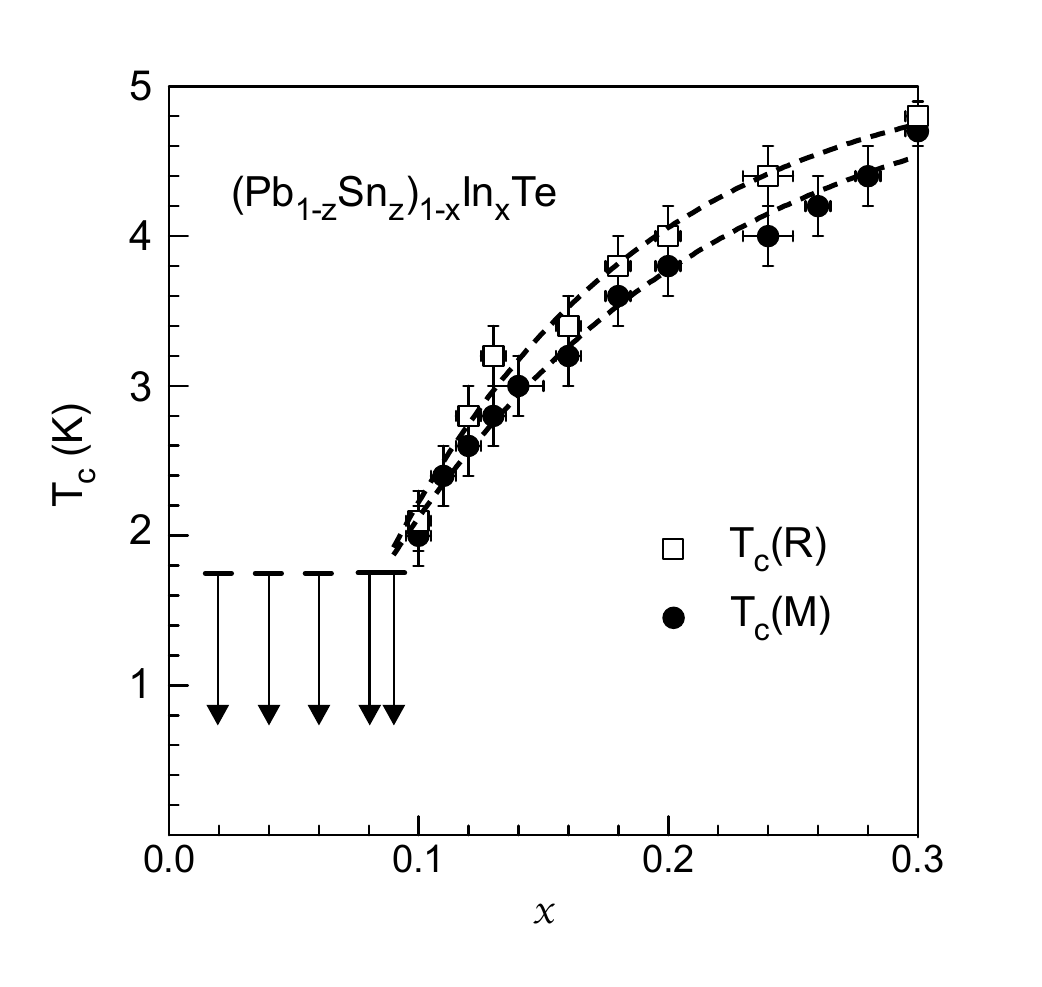}
\caption{\label{fig:phasediagram}Superconducting transition temperature as a function of indium concentration $x$ for \psit\ single crystal samples, obtained from magnetization (solid circles) and resistivity (open squares) measurements. Horizontal error bars reflect the variation of indium concentration measured by EDS. The dashed lines are guides to the eye. Several samples with $x\leq 0.1$ were obtained, but any superconducting critical temperatures were lower than the instrument limit of 1.75~K, represented by arrows.}
\end{center}
\end{figure}

In order to identify the room-temperature crystal structures in this system, single crystal samples with known composition were ground to fine powder and investigated by x-ray diffraction, using Cu $K\alpha$ radiation from a model Rigaku Ultima III, located at the CFN. Figure~\hyperref[fig:XRD]{2(a)} presents the measured XRD patterns, with the intensity plotted on a logarithmic scale.  As can be seen, the peaks of the samples with $0\le x\le0.30$ can be indexed well to the rock-salt structure (space group $Fm\bar{3}m$).  The tetragonal InTe impurity phase only becomes substantial for $x=0.35$, which is consistent with our SEM and EDS results.  Thus, for our single-crystal samples, we estimate that the solubility limit of In in this alloy is approximately 0.30, which is about 50\%\ higher than the limit found in samples prepared by the conventional metal-ceramic technique.\cite{Parfeniev1999, Andrianov2010, Shamshur2009}  In these solid solutions, Pb/Sn and Te form two separate fcc sublattices, with successive substitution of Pb/Sn by In. Figure~\hyperref[fig:XRD]{2(b)} shows the Gaussian fitted (220) peaks, illustrating that the peak position shifts gradually to larger angle before reaching the solubility limit. The $x$ dependence of the corresponding lattice constant $a$ is displayed in Fig.~\hyperref[fig:XRD]{2(c)}. We find that $a=6.392(1)$~\AA\ for $x=0$, decreasing linearly to $a=6.359(2)$~\AA\ for the $x=0.30$ sample. 

\begin{figure}[t]
\begin{center}
\includegraphics[width=0.8\columnwidth]{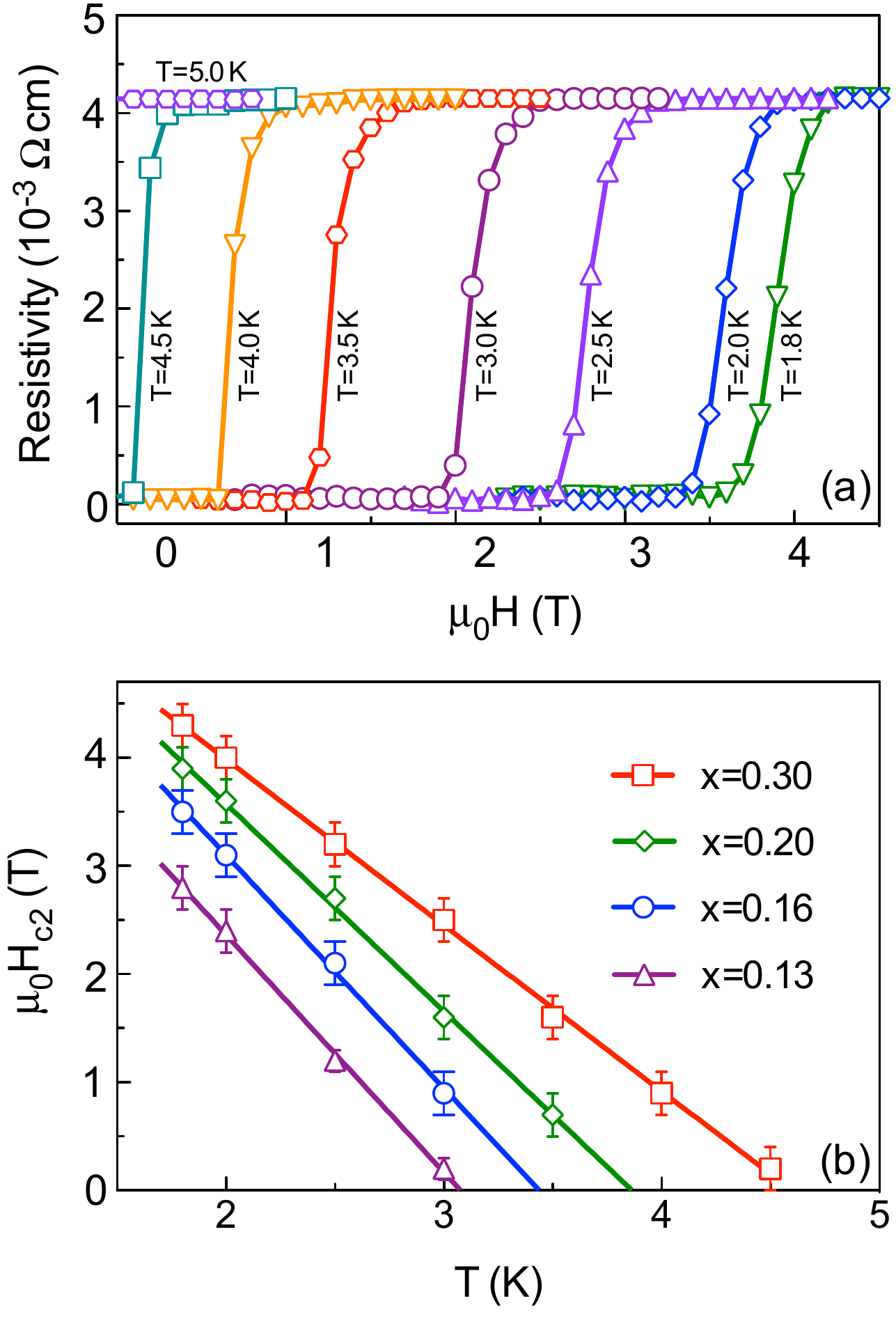}
\caption{\label{fig:Hc2} (Color online) (a) Field dependence of the resistivity for the $x=0.30$ crystal at  fixed temperatures from 1.8 to 5~K. (b) Upper critical field $H_{c2}(T)$ determined from resistivity measurements on four single crystals.}
\end{center}
\end{figure}

To study the effect of the indium substitution on the superconducting properties, we performed both magnetization and resistivity measurements.  The dc magnetic susceptibility measurements were performed using a commercial superconducting quantum interference device (SQUID) magnetometer (MPMS, Quantum Design), for temperatures down to 1.75~K. Electrical resistivity was measured using the standard four-probe configuration, performed by a Keithley digital multimeter (model 2001), with the MPMS used for temperature control.  

Before discussing the superconducting transition, we first consider the striking, non-monotonic variation in the normal-state resistivity with $x$, as illustrated in figure~\hyperref[fig:MTRT]{3(a)}.  For the sample with no In, we find a weakly metallic resistivity.  Increasing $x$ from 0 to 0.06, we observe that the resistivity at 10 K rises by three orders of magnitude.  With further increases of $x$, the resistivity drops, but remains semiconducting, consistent with earlier studies.\cite{Dixon1968, Vul1978, Kozub2006, Shamshur2008}  The initial appearance of superconductivity and the increase in $T_c$ appear to be anti-correlated with the magnitude of the resistivity.  We note that the semiconducting resistivity behavior in the normal state of the superconducting sample is quite different from the case of In-substituted SnTe,\cite{Zhong2013} where all single-crystal samples are weakly metallic in the normal state. 

The critical temperatures $T_{c}(M)$ and $T_{c}(R)$ of each sample were defined as the onset point of the sharp transition of the magnetic moment and resistivity, respectively. Figure~\hyperref[fig:MTRT]{3(b)} shows the temperature dependence of zero-field-cooled (ZFC) and field-cooled (FC) magnetization under an applied field of 1 mT for single crystal sample (Pb$_{0.35}$Sn$_{0.35}$)In$_{0.3}$Te. The onset of the magnetic transition occurs at $T_c=4.7$~K, with a transition width of about 0.5~K (obtained from the temperatures corresponding to 10\%\ and 90\%\ of the full diamagnetism). This rather sharp transition indicates that the indium content is fairly homogeneously distributed in the measured sample.   From the resistivity measurements, we see that only the samples with $x\gtrsim0.10$ show superconductivity at accessible temperatures.  

Superconducting transition temperatures \Tc\ for all samples, obtained from magnetization and resistivity measurements, are summarized in Fig.~\hyperref[fig:phasediagram]{4}. The variation of \Tc\ with $x$ displays a similar relation as for In-substituted SnTe\cite{Zhong2013}: with indium successively doped into the system, the superconducting critical temperature is enhanced. In the \psit\ solid solutions, a maximum in \Tc\ at 4.7~K is achieved in both the magnetic and resistivity measurements. 

To determine the upper critical field, the magnetic-field dependence of the electrical resistivity was also investigated. Representative data for Pb$_{0.35}$Sn$_{0.35}$In$_{0.3}$Te are shown in Fig.~\hyperref[fig:Hc2]{5(a)}. The onset of the resistive transition is plotted as a function of field in Fig.~\hyperref[fig:Hc2]{5(b)} for \psit\ single crystal samples with $x$=0.13, 0.16, 0.20 and 0.30.  From these measures of $H_{c2}(T)$, we have determined the derivative, $|\partial H_{c2}/\partial T|_{T=T_c}$, as indicated by the slopes of the fitted straight lines.  These dependences provide estimates of the critical magnetic fields at zero temperature, $H_{c2}(T=0)$, using the Werthamer-Helfand-Hohenberg approximation,\cite{Werthamer1966}  $H_{c2}(0)=0.69 T_c |\partial H_{c2}/\partial T|_{T=T_c}$. Calculated values for the four superconducting single-crystal samples are compared in Table~\ref{tab:Hc2}.  We find that $\mu_0H_{c2}(0)$ changes relatively little in comparison with $T_c$ for the superconducting samples.

\begin{table}
\caption{Absolute value of the derivative of the upper critical field with respect to the temperature extrapolated to \Tc\ and upper critical fields at zero temperature $\mu_{0} H_{c2}(T=0)$.}
	\begin{tabularx} {0.5\textwidth} {@{}YYYY@{}}
		\toprule         
		$x$ & \Tc & $|\partial\mu_{0}H_{c2}/\partial T|_{T_c}$ & $\mu_{0}H_{c2}(0)$ \\
		 & (K) & (T/K) & (T) \\
		\colrule
		0.13 & 3.2 &  2.20  & 4.86(7)\\
		0.16 & 3.4 & 2.16  &  5.07(8)\\
		0.20 & 4.0 &  1.85  &  5.11(4) \\
               0.30 & 4.8 &  1.54  &  5.10(3) \\
		\botrule
	\end{tabularx}
	\label{tab:Hc2}
\end{table}

To conclude, we have studied the correlation between indium content, crystal structure and superconducting properties for single crystals of \psit.   The solubility limit of In in this system is found to be $x\approx0.30$.  For $x\le0.30$, the crystal remains in the rocksalt structure, with a lattice constant that shrinks linearly with $x$. The superconducting phase diagram, based on the measured In content and the corresponding $T_{c}$(M) and $T_{c}$(R), has been experimentally established.  The highest \Tc\ of 4.7~K was achieved in the crystal of  Pb$_{0.35}$Sn$_{0.35}$In$_{0.3}$Te. This temperature is accessible at liquid helium temperature, and thus provides a good opportunity for testing the topological character of this material and its interaction with the superconductivity.  The upper critical field $\mu_{0}H_{c2}(0)$ of 5~T is fairly substantial considering the modest $T_c$, especially in comparison with Sn$_{1-x}$In$_x$Te.\cite{Zhong2013}

Work at Brookhaven, including at the Center for Functional Nanomaterials, is supported by the U.S. Department of Energy, Office of Science, Office of Basic Energy Sciences, under contract No.\ DE-AC02-98CH10886. R.D.Z. and J.A.S. were supported by the Center for Emergent Superconductivity, an Energy Frontier Research Center funded by the Office of Basic Energy Sciences.

%

\end{document}